\documentclass[acmtog,nonacm]{acmart}
\acmJournal{TOG}
\usepackage{afterpage}
\usepackage[capitalise,noabbrev,nameinlink]{cleveref} % \cref
\usepackage{csquotes} % \enquote
\usepackage{relsize}
\usepackage{mathtools}
\usepackage{multirow,tabularx,arydshln}
\newcolumntype{Y}{>{\centering\arraybackslash}X}
\usepackage{xspace}

\newcommand{\citeetal}[1]{\citeauthor{#1}~\shortcite{#1}}
\newcommand*{\fref}[1]{\hyperref[{#1}]{\ref{#1}~\nameref*{#1}}}
\newcommand{\ImplementationLink}{\href{https://dl.acm.org/journal/tog}{<url anonymised for review>}\xspace}

\renewcommand{\enquote}[1]{“#1”}

\newcommand{\eg}{e.g.\xspace}

\newcommand{\ie}{i.e.\xspace}

\newcommand{\nD}[1]{$#1$D\xspace}         % 1D, 2D, 3D, etc.
\newcommand{\D}{\nD{1}}
\newcommand{\DD}{\nD{2}}
\newcommand{\DDD}{\nD{3}}

\newcommand{\pnorm}[1]{\lVert#1\rVert}
\newcommand{\norm}[1]{\pnorm{#1}_2}

\newcommand{\R}[1]{\mathbb{R}^{#1}}     % R^n
\newcommand{\upbox}[3]{%
    \raisebox{#1}{#3}\kern #2%
}
\newcommand{\UnderPressure}{%
    \textsc{%
        \upbox{-0.51em}{-0.10em}{\larger[2]U\smaller[2]}%
        \upbox{0.0em}{-0.10em}{n}%
        \upbox{0.0em}{-0.08em}{d}%
        \upbox{0.0em}{-0.08em}{e}%
        \upbox{0.0em}{0.0em}{r}%
        \upbox{-0.51em}{-0.08em}{\larger[2]P\smaller[2]}%
        \upbox{0.0em}{-0.11em}{r}%
        \upbox{0.0em}{-0.05em}{e}%
        \upbox{0.0em}{-0.05em}{s}%
        \upbox{0.0em}{-0.08em}{s}%
        \upbox{0.0em}{-0.12em}{u}%
        \upbox{0.0em}{-0.11em}{r}%
        \upbox{0.0em}{0.0em}{e}%
    }\xspace%
}

\begin{document}
    \title{HuMoT: Human Motion Representation using Topology-Agnostic Transformers for Character Animation Retargeting}
    
    \newcommand{\interdigital}{%
    \affiliation{%
      \institution{InterDigital, Inc.}
      \streetaddress{845a, avenue des Champs Blanc}
      \city{Cesson Sévigné}
      \state{Ohio}
      \country{France}
      \postcode{35510}
    }
}

\newcommand{\inria}{%
    \affiliation{%
      \institution{Inria, Univ Rennes, CNRS, IRISA}
      \streetaddress{Campus de Beaulieu Bâtiment 12}
      \city{Rennes}
      \country{France}
      \postcode{35042}
    }
}

\author{Lucas Mourot}
\email{lucas.mourot@interdigital.com}
\orcid{0000-0001-8441-892X}
\interdigital
\inria

\author{Ludovic Hoyet}
\email{ludovic.hoyet@inria.fr}
\orcid{0000-0002-7373-6049}
\inria

\author{François Le Clerc}
\email{francois.leclerc@interdigital.com}
\orcid{0000-0003-0519-8581}
\interdigital

\author{Pierre Hellier}
\email{pierre.hellier@interdigital.com}
\orcid{0000-0003-3603-2381}
\interdigital

    \begin{abstract}
    Motion retargeting is the long-standing problem in character animation that consists in transferring and adapting the motion of a source character to another target character. A typical application is the creation of motion sequences from off-the-shelf motions by transferring them onto new characters. Motion retargeting is also promising to increase interoperability of existing animation systems and motion databases, as they often differ in the structure of the skeleton(s) considered. Moreover, since the goal of motion retargeting is to abstract and transfer motion dynamics, effective solutions might provide expressive and powerful human motion models in which operations such as cleaning or editing are easier. In this article, we present a novel neural network architecture for retargeting that extracts an abstract representation of human motion agnostic to skeleton topology and morphology. Based on transformers, our model is able to encode and decode motion sequences with variable morphology and topology -- extending the current scope of retargeting -- while supporting skeleton topologies not seen during the training phase. More specifically, our model is structured as an autoencoder, and encoding and decoding are separately conditioned on \emph{skeleton templates} to extract and control morphology and topology. Beyond motion retargeting, our model has many applications since our abstract representation is a convenient space to embed motion data from different sources. It may potentially be benefical to a number of data-driven methods, allowing them to combine scarce specialised motion datasets (e.g. with style or contact annotations) and larger general motion datasets, for improved performance and generalisation ability. Moreover, we show that our model can be useful for other applications beyond retargeting, including motion denoising and joint upsampling. Our implementation as well as our pre-trained model can be found at \ImplementationLink.
\end{abstract}
    \begin{CCSXML}
<ccs2012>
   <concept>
       <concept_id>10010147.10010371.10010352.10010380</concept_id>
       <concept_desc>Computing methodologies~Motion processing</concept_desc>
       <concept_significance>500</concept_significance>
       </concept>
   <concept>
       <concept_id>10010147.10010257.10010258.10010260</concept_id>
       <concept_desc>Computing methodologies~Unsupervised learning</concept_desc>
       <concept_significance>500</concept_significance>
       </concept>
   <concept>
       <concept_id>10010147.10010257.10010293.10010294</concept_id>
       <concept_desc>Computing methodologies~Neural networks</concept_desc>
       <concept_significance>500</concept_significance>
       </concept>
 </ccs2012>
 
\ccsdesc[500]{Computing methodologies~Motion processing}
\ccsdesc[500]{Computing methodologies~Unsupervised learning}
\ccsdesc[500]{Computing methodologies~Neural networks}
\end{CCSXML}

    \keywords{
        topology-agnostic,
        retargeting,
        character animation,
        motion representation,
        deep learning,
        transformers,
    }
    

    \maketitle
    
    \begin{figure*}
    \centering
    \includegraphics[width=\textwidth]{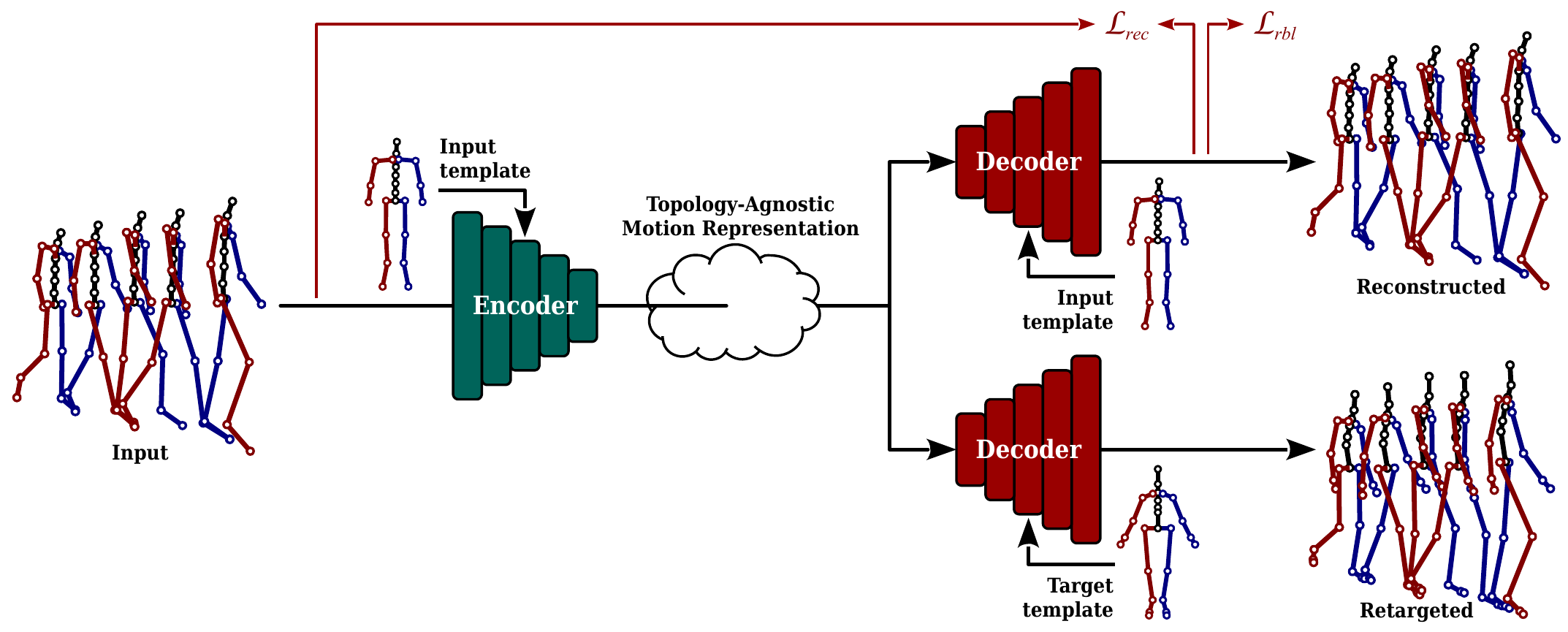}
    \caption{
		Overview of our transformer-based model. It is an autoencoder (\cref{sec:Architecture}) whose encoder and decoder are separetely conditioned on skeleton templates --~which consist in neutral poses (\cref{sec:TemplateEncoding}) -- to control morphological and topological features. During training (upper branch), our model is guided by the reconstruction loss $\mathcal{L}_{rec}$ and the temporal bone lengths consistency loss $\mathcal{L}_{rbl}$ (\cref{sec:Losses}). Applications of our approach include motion retargeting (\cref{sec:EvalMotionRetargeting}), motion denoising (\cref{sec:EvalMotionDenoising}) and joints upsampling (\cref{sec:EvalJointUpsampling}).
    }
    \label{fig:overview}
\end{figure*}
    
    \pagebreak
    \section{Introduction}
        The creation of novel motion sequences has a great importance to character animation. A large spectrum of methods has been proposed, such as the synthesis from high-level parameters (\eg \cite{ref:HoSK16, ref:DHSF19, ref:HYNP20}). Alternatively, the adaptation of existing motion sequences to new contexts (\eg \cite{ref:JaPL22, ref:SCNW19}) can speed up the production of new sequences and avoid costly motion capture. A particular case of motion adaptation is known as motion retargeting and consists in transferring and adapting the motion performed by a source character to a target character with a different skeleton. It is a long-standing problem, which enables the creation of novel motion sequences by transferring off-the-shelf motions onto new characters.

The retargeting task has no formal specification since its goal is to abstract out motion dynamics across different characters: the criterion for success is that the character motion in the retargeted sequence should be the same as in the source. Retargeting was first addressed twenty-five years ago by \citeetal{ref:Glei98}, but recent approaches rely on deep learning. In the general case, the considered source and target characters differ in the lengths of their bones (morphological variations) or in which joints and bones they are composed of (topological variations). However, most existing methods only address morphological variations \cite{ref:VYCL18, ref:KPKH20, ref:LiCC19, ref:LWJZ22}. As an exception, Aberman et al.~\shortcite{ref:ALLS20} recently extended this scope to homeomorphic skeletons only, and their approach does not generalize to unkown skeleton topologies.

In this context, we present a novel approach to motion retargeting by learning an abstract representation of motion that is agnostic to skeleton topology and morphology. Based on transformers, our model extends the scope of retargeting beyond homeomorphic skeletons while supporting skeleton topologies never seen during the training phase. As illustrated in \cref{fig:overview}, it specifically consists in an autoencoder in which both encoder and decoder follow a transformer-based architecture, support motion sequences with variable number of joints, and are separately conditioned on \emph{skeleton templates} to extract and control morphology and topology. Trained on a large amount of motion data gathered from multiple existing databases with different skeleton topologies and morphologies, our model successfully abstracts out motion features from morphological and topological features. Motion retargeting can then be performed through encoding conditioned on the source character followed by decoding conditioned on the target character.

We evaluate our approach on the different cases of retargeting, motion denoising and joint upsampling. Our results demonstrate that it achieves state-of-the-art performance on motion retargeting while extending its scope beyond its current limitations (i.e., limited to homeomorphic skeletons and known topologies). We also show that the topology-agnostic motion representation at the core of our model is well structured as it allows to perform other tasks, such as motion denoising and joint upsampling. Finally, it provides a convenient space to embed motion data from multiple sources with different skeleton morphologies and topologies. For example, data-driven methods could be applied on different data sources, such as scarce specialised motion datasets (\eg with style or contact annotations) together with larger general motion datasets, for improved performance and generalisation ability. Examples illustrating the different results of our approach are provided in the supplementary video.

In summary, our main contributions are the following:
\begin{itemize}
    \item we propose a novel transformer-based architecture applied to motion retargeting, allowing to model complex and distant correlations in joints dynamics,
    \item our method extends the scope of motion retargeting beyond homeomorphic skeletons, toward various human-like skeleton topologies and morphologies, while reaching state-of-the-art performance on homeomorphic retargeting,
    \item we quantitatively evaluate our method on motion retargeting, as well as on side tasks such as motion denoising and joint upsampling.
\end{itemize}
    
    \section{Related Works}
        \subsection{Motion Retargeting}
	Motion retargeting at large might be addressed within different representations of human dynamics. For instance, \citeetal{ref:AWLC19} proposed a \DD~approach based on visual features where inputs and outputs are image sequences. Other variants include the retargeting of body shapes, which considers the dynamics of body surfaces and is notably addressed in parallel to~\cite{ref:VCHY21} or on top of~\cite{ref:MRWF23} skeletal retargeting. Our work is specifically focused on \DDD~skeletal motion retargeting, referred to as retargeting hereafter, which considers motion dynamics of character skeletons as commonly done in character animation.
	
	\citeetal{ref:Glei98} first specifically addressed skeletal motion retargeting. In his work, \citeauthor{ref:Glei98} uses different constraints, \eg on joint locations or body segment orientations, to define specific motion dynamics features that must be preserved or transferred. Retargeting is then performed by numerically solving for the corresponding constrained optimisation problem. Other methods based on iterative optimisation were later proposed \cite{ref:LeSh99, ref:ChKo00, ref:KuMA05, ref:TaKo05, ref:HRED08} but also require hand-crafted kinematic constraints for specific motions~\cite{ref:VYCL18}.
	
    More recently, researchers mostly relied on deep learning to address motion retargeting. Since it is difficult in practice to obtain ground-truth pairs of (source, target) sequences, recent approaches rely on unpaired training data without motion correspondences across characters. As a consequence, these approaches explored a variation of training losses including unsupervised losses, retargeting heuristics and regularisation losses. \citeetal{ref:VYCL18} proposed an encoder-decoder scheme based on recurrent neural networks (RNNs) using an adversarial loss, as well as a cycle consistency loss, encouraging the retargeting to a target character followed by retargeting back to the original character to remain close to the original motion. \citeetal{ref:KPKH20} proposed a similar approach based on a convolutional neural network (CNN), arguing that CNNs are better suited than RNNs for retargeting because they can more accurately capture the short-term motion dependencies that mostly condition the performance of the task. Also based on CNN, \citeetal{ref:LiCC19} simplified the cycle consistency loss into a reconstruction loss applied on self-retargeted motions (\ie same source and target character). Using a model similar to previous approaches in terms of architecture and losses, \citeetal{ref:LWJZ22} additionally proposed to rely on an iterative optimisation at inference time to perform retargeting and encourage source and retargeted motions to have similar normalised end-effector velocities. However, one limitation of all the approaches mentioned above is that only morphological retargeting is possible, \ie source and target characters must have the same skeleton topology.
    
    As an exception, \citeetal{ref:ALLS20} addressed homeomorphic skeleton topologies. For each different topology, a space-time graph-convolutional network (GCN) is trained as an autoencoder to encode to and decode from a shared latent space. Pooling and unpooling operators reduce the different topologies to a common \emph{primal skeleton} used by the latent space. Retargeting is achieved by composing the encoder of the source character with the decoder of the target character. However, the need for a separate autoencoder for each different homeomorphic skeleton topology prevents this approach from easily generalising to unseen topologies.

\subsection{Transformers}
    Transformers refer to a neural network architecture relying on the mechanism of self-attention \cite{ref:VSPU17}, in which the network itself finds out which part of input data it should pay more attention to. First introduced in natural language processing, transformers are gaining popularity in all application domains of deep learning. They have for instance been explored in \DDD facial animation \cite{ref:CZGG22b}. In skeletal character animation, transformer-based models have demonstrated superior performance in various applications such as motion prediction \cite{ref:ACKH20}, style transfer \cite{ref:KMTM22} and motion in-betweening \cite{ref:DSZL21, ref:OVHM22, ref:QiZZ22}.
    
    To the best of our knowledge, we propose the first approach to motion retargeting using self-attention mechanisms implemented in a transformer-based model. This kind of architecture is particularly interesting in the general case of retargeting, \ie with topological and morphological variations, because it allows to find distant and complex correlations in the dynamics of joints through attention mechanisms. This is notably an advantage with respect to RNNs (\eg in \cite{ref:VYCL18}) that tend to focus on the short-term past history, and with respect to GCNs (\eg in \cite{ref:ALLS20}) that mostly capture correlations in the motion of neighboring joints.
	
    \section{Method} \label{sec:Method}
        In contrast to existing retargeting methods, ours gets rid of retargeting heuristics. Indeed, we only rely on a reconstruction loss term to learn a deep representation of motion and a temporal bone lengths consistency loss term to constrain our positional pose representation. Moreover, we leverage attention mechanisms \cite{ref:VSPU17} into an architecture based on transformers, which are a promising alternative to better model distant spatio-temporal correlations. Lastly, our learnt deep motion representation is unified across skeleton morphologies and topologies, pushing the scope of retargeting beyond homeomorphic topologies, and generalises well to novel topologies unseen during training.

Our goal is twofold. First, the architecture of our model needs to be flexible enough to encode variable size inputs to latent variables lying in our deep representation and then to decode them to variable size outputs. In particular, inputs and outputs are motion sequences with variable numbers of frames and joints. Second, we also want our deep representation to be \emph{shared} across skeleton topologies and morphologies. In other words, we want similar motions to be represented by close latent codes in our deep representation, even when skeleton topologies and morphologies differ. To this end, we propose a model relying on an autoencoder (see \cref{fig:overview}) in which both encoder and decoder have transformer-based architectures and are conditioned on \emph{skeleton templates} (see Section~\ref{sec:TemplateEncoding}), representing morphological and topological variations. Our deep motion representation then corresponds to the latent space of our autoencoder and is supposed to encode motion free from skeleton topology and morphology inforamtion. Hence, we make the hypothesis that \enquote{pure} motion features can be reasonably separated from morphological and topological features, in a similar way to shape and pose for faces~\cite{ref:CZGG22a}.

After a brief overview of data representations and notations (\cref{sec:Notation_n_Repr}), we present in detail the proposed model. We begin with the description of the conditioning mechanism used to control and abstract skeleton topology and morphology (\cref{sec:TemplateEncoding}), followed by the proposed architecture (\cref{sec:Architecture}). Finally, we present the procedure used to train our model and how to exploit it at inference time (\cref{sec:Training}).

\subsection{Notation and Representation}\label{sec:Notation_n_Repr}
	\sloppy % prevent formulas overflow in margin
	Our deep representation operates on variable-length motion sequences. We represent them by the position of the joints at each frame, expressed in the global Euclidean space. We note ${X^{(a)} \in \R{\mathcal{J}(a) \times 3 \times F}}$ a motion sequence of $F$~frames \enquote{performed} by the skeleton~$a$ having ${\mathcal{J}(a) \in \mathbb{N}^{+}}$ joints, where the function~$\mathcal{J}(\cdot)$ gives the number of joints of a given skeleton. ${X^{(b)} \in \R{\mathcal{J}(b) \times 3 \times F}}$ denotes a motion sequence consisting of the same motion features as $X^{(a)}$ but performed by another skeleton $b \ne a$. Finally, we note $E$ and $D$ our encoder and decoder modules, ${E(X^{(a)} | \mathcal{T}(a))}$ the encoding of motion sequence~$X^{(a)}$ conditioned on the skeleton template~$\mathcal{T}(a)$ associated to skeleton~$a$, and $D(z | \mathcal{T}(b))$ the decoding of latent code $z$ conditioned on the template~$\mathcal{T}(b)$ associated to skeleton $b$, where the function $\mathcal{T}(\cdot)$ gives the template of given skeleton. We define skeleton templates in the next section.
	
\subsection{Template Conditioning}\label{sec:TemplateEncoding}
	% Introduction
	Our neural network is an autoencoder made up of an encoder $E$ and a decoder $D$, with our deep motion representation corresponding to the latent space lying in between. A first consequence of our goal to abstract out human motion independently of skeleton topology and morphology is the following: no information about topology and morphology should be encoded into the latent space. Therefore, the decoding of such skeleton-agnostic latent codes must be conditioned on the desired topology and morphology of the output motion sequence. To this end, we introduce \emph{skeleton templates} as conditioning representatives of topological and morphological features. In the following, the short form \enquote{template} is equivalent to skeleton template if not stated otherwise.
	
	\begin{figure}
    \centering
    \includegraphics[width=\linewidth]{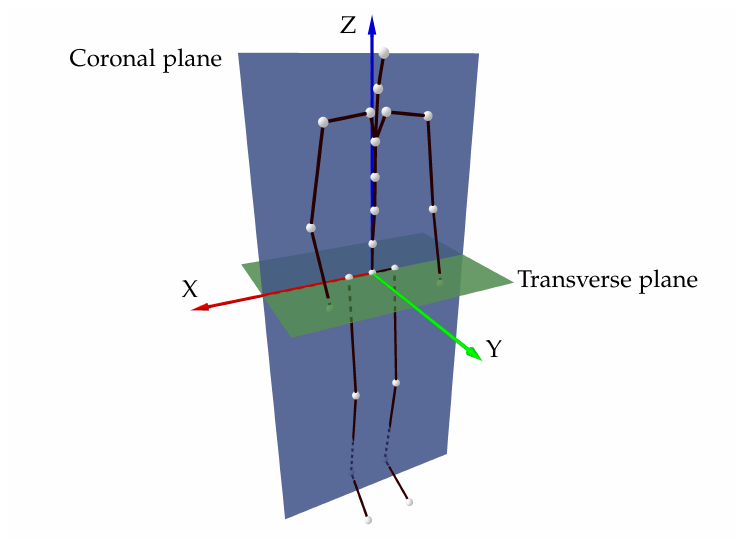}
    \caption{
		Illustration of a typical \emph{skeleton template} used to condition our model, consisting in a neutral standing pose with arms along the body (\ie N-pose), pelvis at the origin of the coordinate system, and transverse (green) and coronal (blue) planes aligned with XY and XZ planes, respectively.
    }
    \label{fig:anatomy}
\end{figure}
	
	% Representation
	In our approach, a skeleton template is a static \emph{neutral} pose, \ie standing with arms along the body (sometimes referred to as N-pose in animation, in opposition to T-pose and A-pose), which we represent by the position of its joints expressed in the global Euclidean space. We normalise the templates such that the position of the pelvis is at the origin, the transverse\footnote{Anatomical plane that divides the body into superior and inferior sections; see \cref{fig:anatomy}.} plane is parallel to the XY-plane and the coronal\footnote{Anatomical plane that divides the body into dorsal and ventral sections; see \cref{fig:anatomy}.} plane is parallel to the XZ-plane. See \cref{fig:anatomy} for a typical skeleton template aligned with transverse and coronal planes. These skeleton templates explicitly encode topological features through the order and position of their joints while morphological features are implicitly encoded through the relative joint positions that correspond to the bone lengths. These poses are required to be neutral for our network to be able to compare templates and to register motion sequences with respect to their respective skeleton templates.
	
	% How to get skeleton templates
	Such static neutral poses are straightforward to obtain. Indeed, topological variations generally appear across animation systems or motion databases, while morphological variations are more common and appear between any different characters or persons. Therefore, after manually predefining a single neutral pose per database or animation system considered, the skeleton template corresponding to a given motion sequence is obtained by scaling the bone lengths of the predefined neutral pose according to the bone lengths observed in the motion sequence. Moreover, the simplicity of neutral poses makes their manual editing straightforward.
	
	% Encoder conditioning and link to next section
	Finally, in addition to our decoder, we condition our encoder on the very same skeleton templates which are expected to be consistent with input motion sequences to be encoded. Still, both encoder and decoder conditionings are independent and different templates might be used to encode and then decode a given motion sequence, \eg to transfer a motion from one character to another, \ie to perform retargeting. As we will see in the following, these templates additionally serve as spatial references when encoding or decoding motion sequences, as a spatial variant of positional encoding which is critical to transformers performance.
	
\subsection{Network Architecture}\label{sec:Architecture}
	Our model relies on an autoencoder network based on the recent transformer models and whose latent space constitutes the proposed unified deep motion representation. This section describes its architecture, while implementation details necessary for reproducibility are provided in \cref{sec:Appendix:Architecture}.
	
	\begin{figure}
    \centering
    \includegraphics[width=\linewidth]{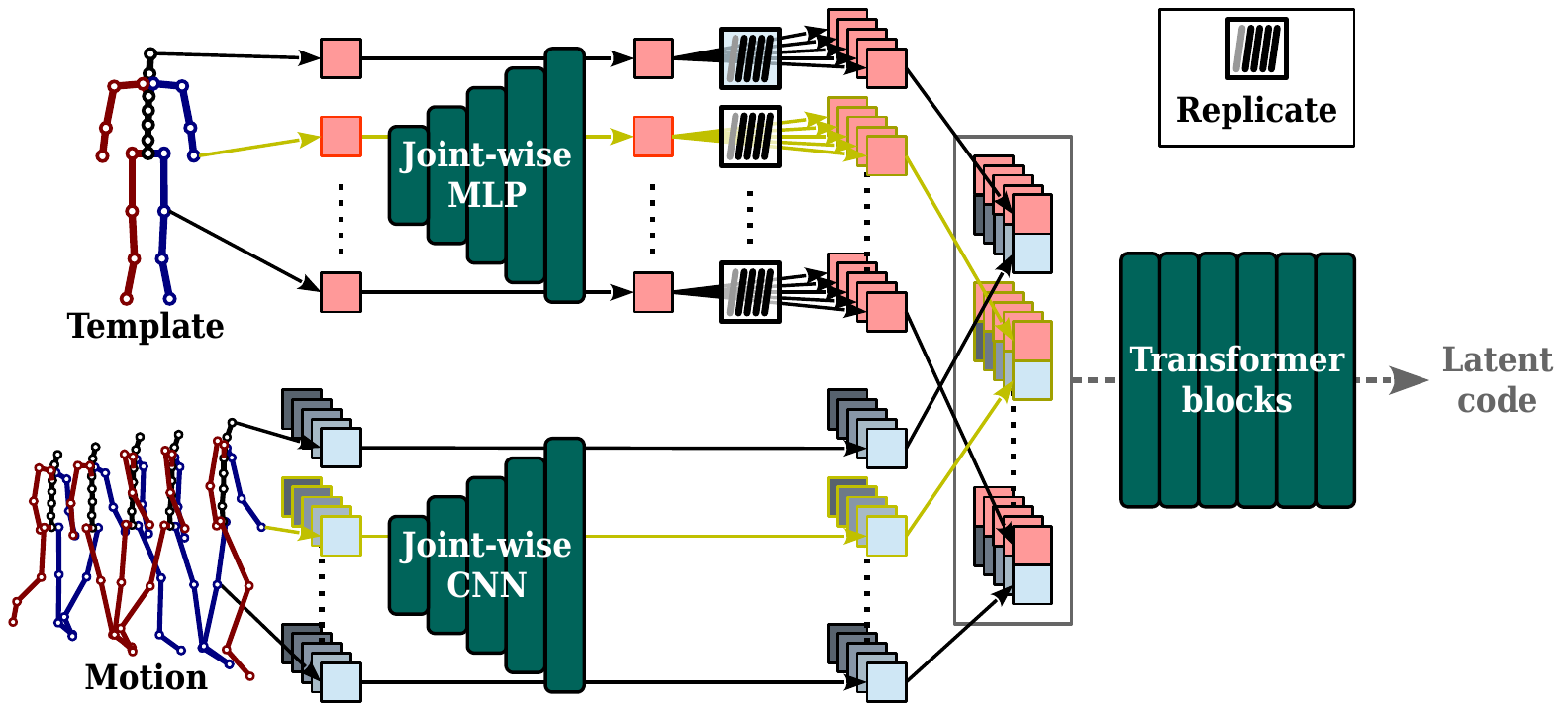}
    \caption{
		Illustration of the spatial positional encoding used in our encoder, performed via joint-wise concatenation of intermediate features of skeleton template and motion. For each joint, motion features are registered to template features of the same joint (\eg left wrist highlight in yellow), telling the network which piece of information is associated to which joint. Thereby, the skeleton template acts as a reference frame for motion features.
    }
    \label{fig:posenc}
\end{figure}
		
	\begin{figure*}
    \centering
    \includegraphics[width=\textwidth]{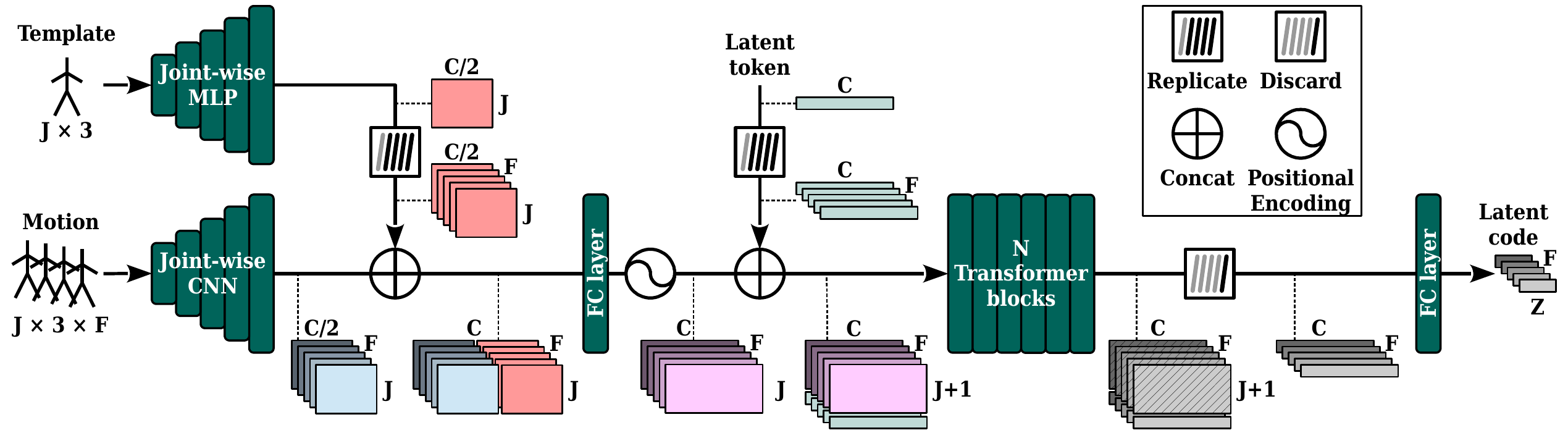}
    \caption{
		Illustration of our encoder architecture: output latent code (right) is computed from input template (top-left) and motion (bottom-left) through the different network layers (dark cyan rounded rectangles) and other functions (see top-right caption), following black arrows. Coloured rectangular slices depict tensor shapes of template (light red), motion (light blue), query token (light cyan) and latent (grey) features at different points in the network. Colour gradients across slices illustrate varying features over time dimension.
    }
    \label{fig:encoder}
    
    \vspace{0.5cm}
    
    \centering
    \includegraphics[width=\textwidth]{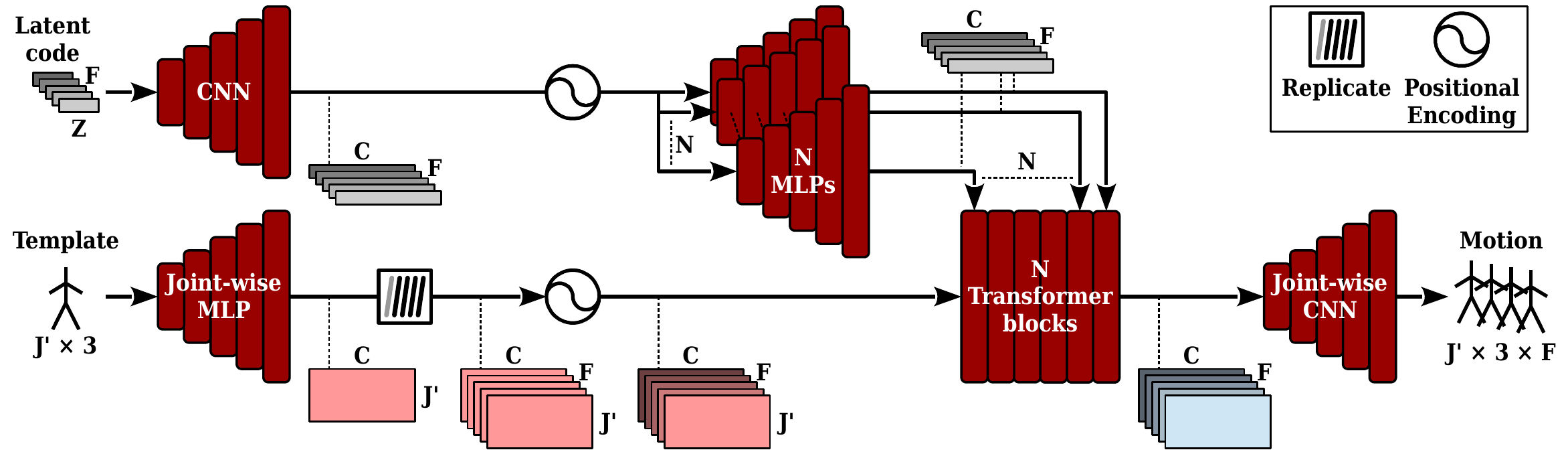}
    \caption{
		Illustration of our decoder architecture: output motion sequence (right) is computed from input template (bottom-left) and latent code (top-left) through the different network layers (dark red rounded rectangles) and other functions (see top-right caption), following black arrows. Coloured rectangular slices depict tensor shapes of template (red), motion (blue) and latent (grey) features at different points in the network. Colour gradients across slices illustrate varying features over time dimension.
    }
    \label{fig:decoder}
    
\end{figure*}
	
	\subsubsection*{Positional Encoding}\label{sec:PositionalEncoding}	
		The inputs and outputs of our model are sequences of joint positions given in an arbitrary order in the spatial domain but consistently across frames. As commonly done in transformer networks, we resort to positional encoding with sine and cosine functions of varying frequencies to give both encoder and decoder knowledge about each frame position in the sequence and hence exploit the temporal structure of motion sequences. To further give information about the spatial structure of input sequences, we rely on another type of positional encoding which consists in concatenating reference features to input features. In our case, this corresponds to concatenating skeleton template features and joint position features, joint by joint, as illustrated in \cref{fig:posenc}. In other words, each joint position of an input motion sequence is registered with the same joint in input skeleton template, hence this mechanism is closely related to our conditioning scheme and to the representation of skeleton templates. Previous work in \DDD shape modelling has shown the benefits of such positional encoding~\cite{ref:CZGG22a, ref:LiWL21}.
	
	\subsubsection*{Encoder}\label{sec:Encoder}
		As illustrated in \cref{fig:encoder}, our encoder is structured as follows: it takes a motion sequence and the corresponding skeleton template as inputs and embeds both of them in respective intermediate representations. The results are merged and fed to a stack of transformer blocks. Both motion sequences and skeleton templates embeddings are performed joint-wise by a \D temporal CNN and a multilayer perception (MLP), respectively. Template embedding is temporally replicated to be concatenated to motion embedding, joint by joint. The result is further processed by a linear layer and modulated by temporal positional encoding to form input tokens of the subsequent transformer blocks.

		As previously stated, our model encodes a given motion sequence~$X^{(a)}$ conditioned on skeleton template~$\mathcal{T}(a)$ to ${z = E(X^{(a)} | \mathcal{T}(a))}$ and decodes it conditioned on another template~$\mathcal{T}(b)$ toward ${\hat{X}^{(b)} = D(z | \mathcal{T}(b))}$. In such a scenario, both $X^{(a)}$ and $\hat{X}^{(b)}$ will have the same number of frames, but might have a different number of joints. Our model retains the sequence-to-sequence paradigm of transformer networks along the temporal dimension. Along the spatial dimension the number of joints may differ across inputs and outputs. Since our goal is to produce a latent code that is agnostic to the skeletal structure, we impose a latent code of fixed spatial dimension. Thus, in the spatial domain our encoder and decoder act as sequence-to-item and item-to-sequence functions, respectively.
		
		To achieve this, we use an additional learned token, called \emph{latent token}, which is appended to hidden features. First introduced in BERT \cite{ref:DCLT19}, this additional sequence-level token, sometimes referred to as \texttt{[CLS]}, is a common design pattern in transformer architectures (\eg \cite{ref:CZGG22a}). In our case, it is used as a fixed-size container for output features while other output tokens of the final transformer block are discarded. During training, the network is thus forced to progressively migrate necessary information toward the latent token along transformer blocks.
		
		To obtain latent codes with fixed spatial size but variable temporal size, we temporally replicate the latent token to match the number of frames of the input motion before concatenating it to hidden features. More formally, the latent token is a tensor of shape~${1 \times C}$, where $C$ is a number of intermediate features. Given an input motion sequence~${X^{(a)} \in \R{\mathcal{J}(a) \times 3 \times F}}$, the sequence of tokens obtained after positional encoding will be a tensor of shape~${\mathcal{J}(a) \times C \times F}$. Then, the latent token is replicated $F$~times, producing a tensor of shape~${1 \times C \times F}$ and appended to the sequence of tokens, resulting in a tensor of shape~${(\mathcal{J}(a)+1) \times C \times F}$, which will be the actual input sequence to the transformer blocks. Finally, at the end of the final transformer block, the first~$\mathcal{J}(a)$~slices are discarded to get a tensor of shape~${1 \times C \times F}$ which is further processed by a linear layer toward the latent code~${z \in \R{Z \times F}}$ where $Z$ is the spatial size of the latent space. See \cref{fig:encoder} for a better understanding of how these different tensors are handled into our encoder.

	\subsubsection*{Decoder}\label{sec:Decoder}
		As depicted in \cref{fig:decoder}, our decoder is structured as follows: it takes a latent code~${z \in \R{Z \times F}}$ and a skeleton template~${\mathcal{T}(b) \in \R{\mathcal{J}(b) \times 3}}$ as inputs and decodes them toward a motion sequence ${\hat{X}^{(b)} = D(z | \mathcal{T}(b)) \in \R{\mathcal{J}(b) \times 3 \times F}}$ of the same temporal length as the latent code~$z$ but the same number of joints as the skeleton template~$\mathcal{T}(b)$. To do so, the skeleton template~$\mathcal{T}(b)$ is first embedded into an intermediate representation using a joint-wise MLP, then replicated $F$ times to match the length of the latent code. Eventually, temporal positional encoding is applied.
		
		The result is used as input tokens to a stack of transformer blocks, as in our encoder. However, differently from the encoder, information derived from the latent code is injected into each block rather than only as input to the first block. This dissemination of latent information at different points of the network, sometimes called \emph{style modulation}, has proven to be effective when it comes to synthesising new samples, \eg in the \emph{StyleGAN}~series~\cite{ref:KaLA19, ref:KLAH20, ref:KALH21}. In our decoder, each style-modulated transformer block corresponds to the same transformer block as in our encoder with an additional \emph{style} module which maps the latent code into a style code. The style code is then used to modulate input tokens of the transformer block through multiplication. Style modules are \D temporal CNNs, well suited to process latent codes that are temporally structured (see~\hyperref[sec:Encoder]{previous section}). Most of the first layers of style modules are shared across blocks (see \cref{sec:Appendix:Architecture}).

		Finally, the output of the last transformer block is further processed by a joint-wise \D temporal CNN to obtain the final motion sequence.
	
	\subsubsection*{Cross-Covariance Attention}\label{sec:XCA}
		In this work, we consider motion sequences and model interactions between joint positions at any frame. However, one of the limitations of transformers is the time and memory complexity of self-attention layers, which increase quadratically with the number of input tokens \cite{ref:ETCB21}. In our model, we instead leverage cross-covariance attention (XCA), recently proposed by \citeetal{ref:ETCB21}, which is a transposed version of self-attention with linear complexity with respect to the number of tokens. Rather than operating on tokens, and computing interactions based on the cross-covariance matrix between keys and queries, it operates across feature channels. \citeetal{ref:ETCB21} proposed a cross-covariance image transformer (XCiT) block adapted from the original transformer block \cite{ref:VSPU17} with three main changes: first, the self-attention layer is replaced with an XCA~layer; second, a local interaction layer is added in-between attention and feed-forward layers; third, layer normalisation is applied before rather than after each layer of the block. The additional local interaction layer enables explicit communication across tokens (only implicit in XCA~layer) and consists in two \DD convolutional layers. Even though our model operates on motion sequences rather than images, we build it on top of the same cross-covariance transformer blocks. Indeed, the only part that might be sensitive to the type of data processed is the local interaction layer made of \DD convolutional layers. Since its goal is to make tokens interact locally, \DD convolutions are as relevant to image patches as to spatio-temporal motion sequence patches.
        
\subsection{Training}\label{sec:Training}
	\subsubsection{Data Preparation} \label{sec:DataPreparation}
\begin{figure*}
    \centering
	\includegraphics[width=\textwidth]{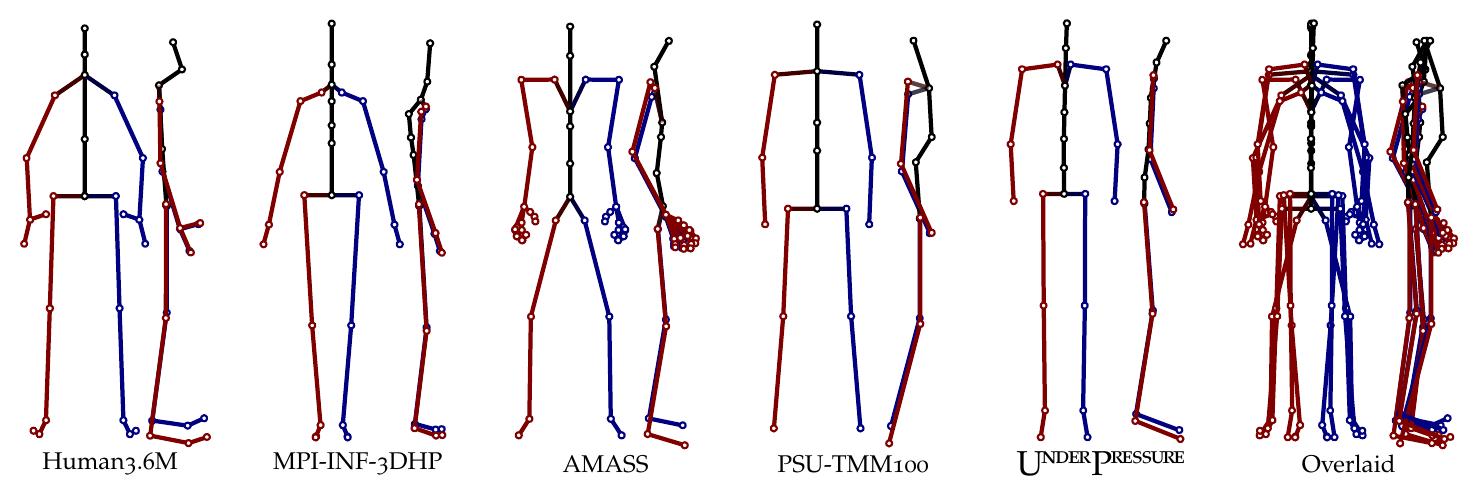}
    \caption{
		Illustration of the different skeleton topologies present in the datasets we gathered, and their structural variations, represented by the corresponding predefined neutral poses. The left-most five columns show front and side views of the different topologies. Each one is associated to a subset of the data (from left to right: Human3.6M~\cite{ref:IPOS14}, MPI-INF-3DHP~\cite{ref:MRCF17}, AMASS~\cite{ref:MGTP19}, PSU-TMM100~\cite{ref:SRFC20}, and \UnderPressure~\cite{ref:MHLH22}). The right-most column gives an insight on topological variations by overlaying the different topologies.
    }
    \label{fig:topologies}
\end{figure*}

		The goal of our approach is to learn the structure of the human body from data instead of relying on a single predefined skeleton topology. To this end, we collected a large amount of motion data gathered from existing public databases (Human3.6M~\cite{ref:IPOS14}, MPI-INF-3DHP~\cite{ref:MRCF17}, AMASS~\cite{ref:MGTP19}, PSU-TMM100~\cite{ref:SRFC20} and \UnderPressure~\cite{ref:MHLH22}) and internal motion data. In total we have more than 65~hours of motion data, spanning 5~different skeleton topologies (see \cref{fig:topologies}) and 494 different morphologies.
		
		\begin{table*}
    \small
    \centering
    \caption{
		Overview of the different datasets gathered to train our model.
    }
    \begin{tabularx}{\textwidth}{ c | YcY | Yc | c }
		Dataset & Frames & Framerate & Size & Joints & Repr. & Availability \\
		\hline		
		
		Human3.6M \cite{ref:IPOS14}      & 3.6×10$^6$ & 50 Hz      & 20.0 h & 25 & angular    & On request \\
		MPI-INF-3DHP \cite{ref:MRCF17}   & 1.3×10$^5$ & 25 Hz      &  1.5 h & 27 & positional & Public \\
		AMASS \cite{ref:MGTP19}          & 1.8×10$^7$ & 60--250 Hz & 41.5 h & 52 & angular    & Public     \\
		PSU-TMM100 \cite{ref:SRFC20}     & 1.3×10$^6$ & 50 Hz      &  7.4 h & 17 & positional & On request \\
		\UnderPressure \cite{ref:MHLH22} & 4.9×10$^6$ & 240 Hz     &  5.6 h & 23 & angular    & Public \\	
		
    \end{tabularx}
    \label{tab:datasets}
\end{table*}
		
		Data is preprocessed in three steps:
		\begin{enumerate}
			\item We resample all motion sequences to a common framerate (\ie~30~Hz) using spherical linear interpolation on motion sequences represented with angular pose representations and linear interpolation on motion sequences represented with positional pose representations.		
			\item We convert all motion sequences to joint positions expressed in the same $XYZ$ Euclidean space, by switching their axes, converting their coordinates to meters and/or applying forward kinematics on motion sequences represented with angular pose representations to get joint positions when needed.
			\item We extract 1-second overlapping chunks (\ie 30-frame chunks overlapping over 24 frames, or 0.8 s) and normalise them separately by removing the mean position computed over a set of major joints (\ie ankles, hips, knees, shoulders, elbows, wrists).
		\end{enumerate}
		The resulting unified, augmented and normalised motion sequences constitute our dataset, for a total of about 300 hours of motion data.
	
	\subsubsection{Training Objective} \label{sec:Losses}
	    Similarly to a typical autoencoder, our model is trained by minimising errors in outputs with respect to inputs (reconstruction loss). However, we slightly change this scheme to further prevent our model from being tied to one or more particular skeleton topology, and rely on a \emph{stochastic joint subsampling} of input and target motion sequences. Moreover, we use a temporal consistency loss over bone lengths. Both objectives are described in the following paragraphs.
		
		\paragraph{Reconstruction loss.}
			The idea of stochastic joint subsampling comes by thinking of motion sequences tied to their skeleton topologies as proxy representations of their underlying motion features. Within this context, we can discard a few joints from motion sequences and still consider their underlying motion features to be the same. Then, given an input motion sequence~$X^{(a)}$ sampled from the training set, we first derive two sub-versions ${X^{(a_1)} = \mathcal{S}(X^{(a)})}$ and ${X^{(a_2)} = \mathcal{S}(X^{(a)})}$, where~$\mathcal{S}(\cdot)$ stands for stochastic joint subsampling and discards a random set of joints. Skeleton template $\mathcal{T}(a)$ is subsampled consistently with $X^{(a_1)}$ and $X^{(a_2)}$ subsamplings to get templates ${\mathcal{T}(a_1) = \mathcal{S}(\mathcal{T}(a))}$ and ${\mathcal{T}(a_2) = \mathcal{S}(T(a))}$, respectively. We consider these two sub-versions as different proxy representations of the same underlying motion features $X$. We encode~$X^{(a_1)}$ and then decode the resulting latent code~${z = E(X^{(a_1)} | \mathcal{T}(a_1))}$ conditioned on $a_2$ to get ${\hat{X}^{(a_2)} = D(z | \mathcal{T}(a_2))}$. Finally, our reconstruction loss is implemented to minimise the discrepancy between $\hat{X}^{(a_2)}$ and $X^{(a_2)}$. This mechanism is critical to the training of our model as it forces both encoder and decoder to strictly follow the conditioning template skeletons they are given to encode and decode motion sequences. Indeed, the variations in skeleton topologies seen by our network is much larger than in existing retargeting approaches. Formally, we compute the deviation of $\hat{X}^{(a_2)}$ with respect to $X^{(a_2)}$ as the mean squared position error:
			\begin{equation} \label{eq:loss:rec}
				\mathcal{L}_{rec} = \frac{1}{F} \frac{1}{\mathcal{J}(a_2)} \sum_{f=1}^{F}{\sum_{j=1}^{\mathcal{J}(a_2)}{%
					\norm{X_{f,j}^{(a_2)} - \hat{X}_{f,j}^{(a_2)}}^2
				}}
			\end{equation}
		
        \begin{table*}
    \small
    \centering
    \caption{
		Quantitative evaluation of the representation accuracy of our deep motion representation. We measure the distortion introduced when encoding and then decoding ground truth motion sequences drawn from our validation set with the MPJPE in centimetres. We distinguish skeleton topologies that have been \emph{seen} during training from \emph{unseen} topologies. Columns are associated to the different skeleton topologies in our validation set, named after the data sources used to constitute our dataset (see \cref{sec:Appendix:Data}), while rows correspond to the variants of our approach (see \cref{sec:DataPreparation}). Lower MPJPE values mean higher representation accuracy.
    }
    \begingroup
    \begin{tabular}{ c | ccccc | c }
        \multirow{2}{*}{Model} & \multicolumn{5}{c|}{Seen} & Unseen \\
        \cdashline{2-7}[0.3mm/0.3mm]
        & Human3.6M & AMASS & PSU-TMM100 & \UnderPressure &  \multicolumn{1}{|c|}{Overall} & MPI-INF-3DHP \\
        \hline
    % 	Ours - no SJS &
    %     	2.89 ± 0.30 cm &
    %     	1.95 ± 0.41 cm &
    %     	3.03 ± 0.55 cm &
    %     	2.93 ± 0.21 cm &
    %     	\multicolumn{1}{|c|}{2.19 ± 0.58 cm} &
    %     	4.22 ± 1.28 cm \\
        
    % 	Ours - no BLC &
    %     	2.78 ± 0.50 cm &
    %     	1.93 ± 0.67 cm &
    %         2.05 ± 0.21 cm &
    %     	2.08 ± 0.38 cm &
    %     	\multicolumn{1}{|c|}{1.99 ± 0.65 cm} &
    %     	4.07 ± 0.89 cm \\ % as of 2023-05-20
        
    % 	Ours &
    % 	    \textbf{1.81 ± 0.19 cm} &
    % 	    \textbf{1.01 ± 0.32 cm} &
    % 	    \textbf{1.75 ± 0.48 cm} &
    % 	    \textbf{1.22 ± 0.20 cm} &
    % 	    \multicolumn{1}{|c|}{\textbf{1.13 ± 0.41 cm}} &
    % 	    \textbf{3.09 ± 0.86 cm} \\
        
        Ours - no SJS &
            3.03 ± 0.55 cm &
            1.95 ± 0.41 cm &
            2.89 ± 0.30 cm &
            2.93 ± 0.21 cm &
            \multicolumn{1}{|c|}{2.19 ± 0.58 cm} &
            4.22 ± 1.28 cm \\
        
        Ours - no BLC &
            2.48 ± 0.45 cm &
            1.97 ± 0.65 cm &
            \textbf{1.81 ± 0.19} cm &
            1.97 ± 0.34 cm &
            \multicolumn{1}{|c|}{1.97 ± 0.60 cm} &
            3.75 ± 0.81 cm \\
        
        Ours &
            \textbf{1.75 ± 0.48 cm} &
            \textbf{1.01 ± 0.32 cm} &
            \textbf{1.81 ± 0.19 cm} &
            \textbf{1.22 ± 0.20 cm} &
            \multicolumn{1}{|c|}{\textbf{1.13 ± 0.41 cm}} &
            \textbf{3.09 ± 0.86 cm} \\

    \end{tabular}
    \endgroup
    \label{tab:reconstruction}
\end{table*}

		\paragraph{Temporal Bone Lengths Consistency.}
		    As explained in \cite{ref:MHLS21}, positional pose representations do not constrain bone lengths to remain constant over time, which can be problematic when synthesising motion sequences. A common approach to solve this issue consists in adding some constraint to ensure that distances between pairs of adjacent joints, referred to as bones, are consistent~\cite{ref:LZZX19, ref:LZZL20}. Therefore, we rely on an additional temporal bone lengths consistency loss to train our model such that the motion sequences produced by the decoder have consistent bone lengths over time. It consists in minimising the temporal variance of bone lengths in the decoder output. However, bone lengths have large variations across the different bones and thus minimising the absolute temporal variance of bone lengths might fail to constrain short bones. We instead use the relative temporal variance of bone lengths obtained by normalising, per bone, output lengths with respect to the ground truth lengths. Our temporal bone lengths consistency loss is then:
			\begin{equation} \label{eq:loss:blc}
				\mathcal{L}_{blc} = \frac{1}{\mathcal{B}(a_2)} \sum_{b=1}^{\mathcal{B}(a_2)}{%
					Var \left( \frac{
						\hat{X}_b^{(a_2)}
					}{
						X_b^{(a_2)}
					} \right)
				}
			\end{equation}
			where function $\mathcal{B}(\cdot)$ gives the number of bones of a given skeleton and ${X_b^{(a)} \in \R{F}}$ denotes the vector of lengths of the $b$th~bone at each of the $F$~frames of motion sequence~$X^{(a)}$.
		
		In summary, we train our model with two objectives, a reconstruction loss and a temporal bone lengths consistency loss. The total loss used to train our model is ${\mathcal{L} = \mathcal{L}_{rec} + \lambda_{blc} \cdot \mathcal{L}_{blc}}$, where $\lambda_{blc}$ is a hyperparameter to balance both losses.

    \section{Evaluation} \label{sec:Eval}
        In this section, we evaluate the representation accuracy of our model (\cref{sec:EvalRepresentationAccuracy}), as well as its performance on motion retargeting (\cref{sec:EvalMotionRetargeting}). Then, we further show that our deep motion representation is well structured through the evaluation of additional tasks, namely motion denoising (\cref{sec:EvalMotionDenoising}) and joint upsampling (\cref{sec:EvalJointUpsampling}). Animation examples that are meaningful for evaluating the quality of our approach are also provided in the supplementary video. Please refer to \cref{sec:Appendix} for implementation details.

Throughout our evaluations, we perform an ablative study of our model to show the relevance of the bone length consistency~(\emph{BLC}) term in our loss, as well as of our stochastic joint subsampling mechanism (\emph{SJS}, see \cref{sec:Losses}). Hereafter, we note \emph{Ours~-~no~BLC} and \emph{Ours~-~no~SJS} the variants of our main model (called \emph{Ours}) that were trained without the bone length consistency and without the stochastic joint subsampling mechanism, respectively.
		
\subsection{Representation Accuracy}\label{sec:EvalRepresentationAccuracy}
	A key aspect of deep representations is their ability to accurately encode information, sometimes called the representation accuracy. In this section, we evaluate this aspect by quantifying the amount of distortion introduced by our deep motion representation when encoding and then decoding motion sequences drawn from our validation set (see \cref{sec:Appendix:Data}). We measure this amount of distortion using the mean per joint position error (MPJPE) of reconstructed motion sequences with respect to the motion sequences at the encoder input.
	
	\cref{tab:reconstruction} provides the corresponding results for both our main model and its variant without stochastic joint subsampling and for each skeleton topology. We separate the evaluation between skeleton topologies that have been seen during training from the unseen skeleton topologies, \ie the ones associated to MPI-INF-3DHP. Overall our model performs well with an average reconstruction error of 1.13 cm on skeleton topologies seen during training and 3.09 cm on the topology of MPI-INF-3DHP, never seen by our network. For example, the order of magnitude of the reconstruction of the pose autoencoder proposed by \citeetal{ref:ViMB21} for edition is about 1 cm (computed on a single sample pose). Moreover, the ablative study shows the importance of both stochastic joint subsampling and the bone length consistency loss, without which the representation accuracy is significantly degraded.
	
\begin{figure*}
	\centering
	\includegraphics[width=\textwidth]{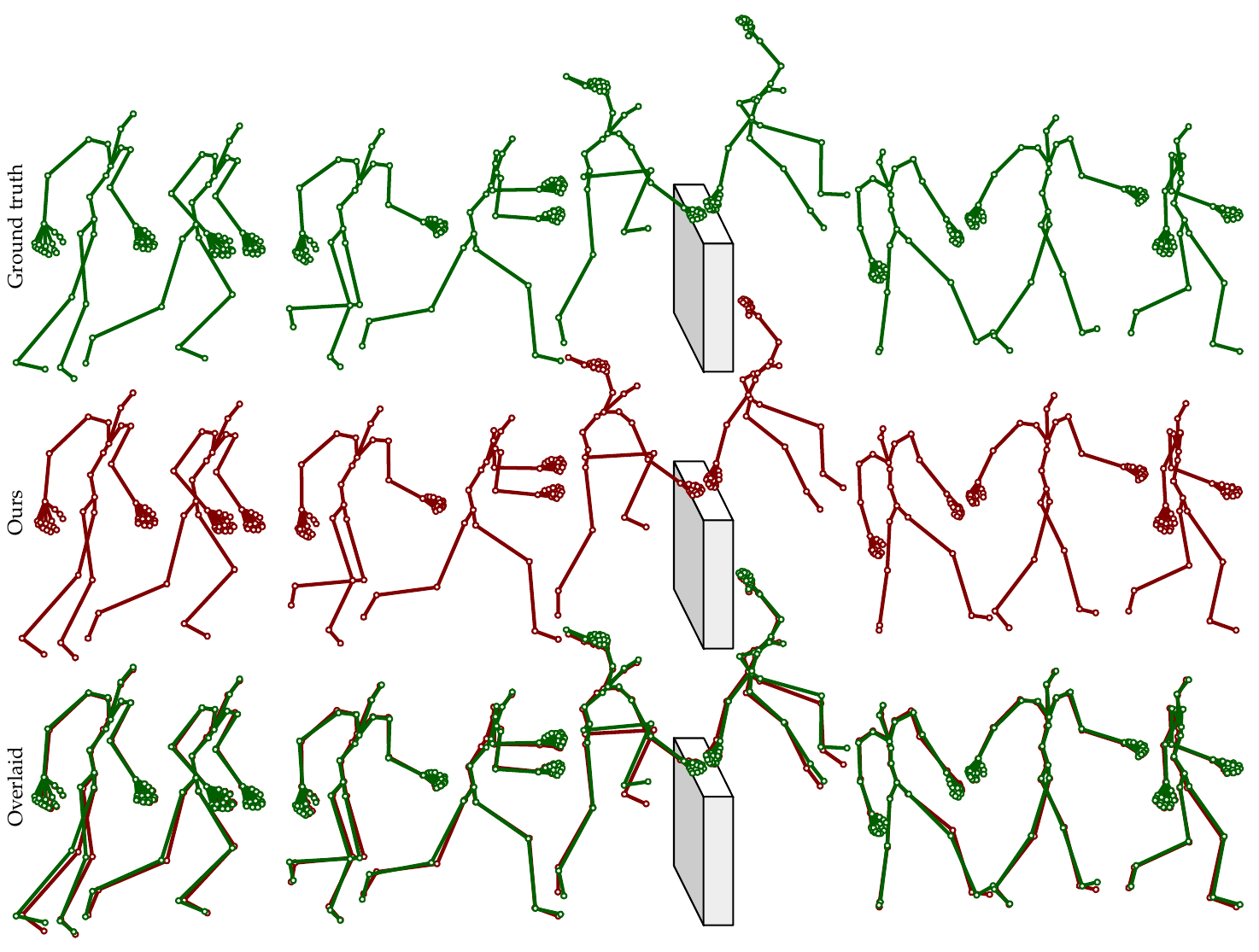}
    \caption{
		Illustration of the representation accuracy of our deep motion representation on a challenging motion sequence, consisting in a speed vault to cross some fence or wall, depicted in white. The ground truth test motion sequence (top row, green) is given as input to our model which encodes and then decodes it to get the corresponding reconstructed motion sequence (middle row, red). The bottom row shows both ground truth and reconstructed motion sequences overlaid to emphasise reconstruction errors.
    }
    \label{fig:reconstruction}
\end{figure*}

	\cref{fig:reconstruction} also illustrates the representation accuracy of our model on a challenging  motion sequence. In this example, we can see that the dynamics is relatively well encoded with fine details, such as hand orientations which are kept consistent throughout the sequence. Positional errors mostly occur during fast and ample movements such as leg swings when the character is running or jumping over the fence. Still, the MPJPE amounts to 1.33~cm on this example, which is only slightly above global and topology averages (1.13~cm and 1~cm, respectively; see \cref{tab:reconstruction}).
    
\subsection{Motion Retargeting}\label{sec:EvalMotionRetargeting}
	As pointed out by \citeetal{ref:ALLS20}, the task of motion retargeting has no formal specification as its goal is to abstract out motion dynamics. Nevertheless, research in motion retargeting relied on the Mixamo dataset \cite{ref:MI21} as a source of motion sequences performed by different characters, considered as ground truth for evaluation purposes \cite{ref:VYCL18, ref:LiCC19, ref:ALLS20, ref:KPKH20, ref:LWJZ22}. In this section, we provide quantitative retargeting results following the evaluation procedure from \citeetal{ref:ALLS20}, as well as qualitative visual results. Animated results can also be found in the supplementary video. 
    
    We use the same 106 test motion sequences from the Mixamo dataset as \citeetal{ref:ALLS20} and divide the evaluation into two modes, called intra-structural and cross-structural. The former corresponds to motion retargeting with the same skeleton topology but different body proportions, while the latter additionally considers source characters with a topology different from the target. Five characters are considered: \emph{BigVegas}, \emph{Goblin}, \emph{Mousey}, \emph{Mremireh} and \emph{Vampire}, respectively noted $B$, $G$, $M_o$, $M_r$ and $V$ hereafter. \emph{BigVegas} has a different topology from the other characters and is used as the source character to evaluate cross-structural retargeting. For intra-structural retargeting, \citeetal{ref:ALLS20} considered all combinations of the last four characters as source and target characters. However, according to their publicly released implementation\footnote{\href{https://github.com/DeepMotionEditing/deep-motion-editing}{https://github.com/DeepMotionEditing/deep-motion-editing}\label{SAN_code}}, these combinations are drawn with replacement, meaning that their evaluation of intra-structural retargeting includes self-retargeting, \ie retargeting with identical source and target characters. In the following, we further evaluate self-retargeting apart from intra-structural retargeting.
    % Then to measure the retargeting error, we use the MPJPE normalised by the height of the skeleton.
    \citeetal{ref:ALLS20} measure the retargeting error with the squared error\footnotemark[\getrefnumber{SAN_code}] normalised by the height of the skeleton. Although the squared error is generally well suited as a loss function, it is less meaningful for evaluation. For this reason, we instead rely on the MPJPE normalised by the height of the skeleton.
    
    \begin{table*}
    \small
    \centering
    \caption{
		Quantitative evaluation of self-retargeting, \ie from a source to a target character with the same skeleton topology and body proportions. We measure the retargeting error with the MPJPE normalised by the height of the skeleton, and provide per character (middle) as well as overall (right-most) results. $B$, $G$, $M_o$, $M_r$ and $V$ stand for Big Vegas, Goblin, Mousey, Mremireh and Vampire characters, respectively. Lower values mean higher retargeting accuracy and \textbf{bold} and \underline{underlined} indicate best and second best, respectively.
    }
    \begin{tabular}{ c | cccc | c }
        Model & $G \rightarrow G$ & $M_o \rightarrow M_o$ & $M_r \rightarrow M_r$ & $V \rightarrow V$ & Overall \\
        \hline
        SAN \shortcite{ref:ALLS20} &
            0.0181 ± 0.0099 &
            \textbf{0.0150 ± 0.0080} &
            0.0172 ± 0.0097 &
            0.0210 ± 0.0113 &
            0.0178 ± 0.0100 \\
        Ours - no SJS &
            \textbf{0.0054 ± 0.0009} &
            0.0200 ± 0.0016 &
            \textbf{0.0052 ± 0.0006} &
            \textbf{0.0068 ± 0.0010} &
            \underline{0.0093 ± 0.0063} \\
        Ours - no BLC &
            0.0090 ± 0.0016 &
            0.0192 ± 0.0019 &
            0.0081 ± 0.0015 &
            0.0090 ± 0.0016 &
            0.0113 ± 0.0048 \\
        Ours &
            \underline{0.0063 ± 0.0013} &
            \underline{0.0153 ± 0.0031} &
            \underline{0.0059 ± 0.0011} &
            \underline{0.0071 ± 0.0014} &
            \textbf{0.0087 ± 0.0043} \\

    \end{tabular}
    \label{tab:self}
\end{table*}

    \begin{table*}
    \small
    \centering
    \caption{
		Quantitative evaluation of intra-structural retargeting, \ie from a source to a target character with the same skeleton topology. We measure the retargeting error with the MPJPE normalised by the height of the skeleton, and provide per character (middle) as well as overall (right-most) results. $G$, $M_o$, $M_r$ and $V$ stand for Goblin, Mousey, Mremireh and Vampire characters, respectively. Lower values mean higher retargeting accuracy and \textbf{bold} and \underline{underlined} indicate best and second best, respectively. 
    }
    \begin{tabular}{ c | cccccc | c }
        Model &	
		$G \leftrightarrow M_o$ &
		$G \leftrightarrow M_r$ &
		$G \leftrightarrow V$ &
		$M_o \leftrightarrow M_r$ &
		$M_o \leftrightarrow V$ &
		$M_r \leftrightarrow V$ &
		Overall \\
		
        \hline
        SAN \shortcite{ref:ALLS20} &
            \textbf{0.0266 ± 0.0148} &
            \textbf{0.0326 ± 0.0206} &
            \textbf{0.0275 ± 0.0153} &
            \textbf{0.0360 ± 0.0246} &
            \textbf{0.0358 ± 0.0235} &
            \textbf{0.0240 ± 0.0141} &
            \textbf{0.0304 ± 0.0198} \\
        
        Ours - no SJS &
            0.0783 ± 0.0291 &
            0.0480 ± 0.0078 &
            \underline{0.0533 ± 0.0126} &
            0.0746 ± 0.0296 &
            0.0875 ± 0.0249 &
            0.0412 ± 0.0191 &
            0.0638 ± 0.0279 \\
        
        Ours - no BLC &
            0.0672 ± 0.0290 &
            \underline{0.0510 ± 0.0085} &
            0.0553 ± 0.0120 &
            0.0710 ± 0.0302 &
            0.0760 ± 0.0198 &
            \underline{0.0328 ± 0.0210} &
            0.0589 ± 0.0260 \\
        
        Ours &
            \underline{0.0666 ± 0.0300} &
            0.0551 ± 0.0083 &
            0.0548 ± 0.0126 &
            \underline{0.0649 ± 0.0312} &
            \underline{0.0672 ± 0.0211} &
            0.0396 ± 0.0195 &
            \underline{0.0580 ± 0.0241} \\
    \end{tabular}
    \label{tab:intra}
\end{table*}

    \begin{table*}
    \small
    \centering
    \caption{
		Quantitative evaluation of cross-structural retargeting, \ie from a source (BigVegas here) to target characters with a different skeleton topology. We measure the retargeting error with the MPJPE normalised by the height of the skeleton, and provide per character (middle) as well as overall (right-most) results. $B$, $G$, $M_o$, $M_r$ and $V$ stand for Big Vegas, Goblin, Mousey, Mremireh and Vampire characters, respectively. Lower values mean higher retargeting accuracy and \textbf{bold} and \underline{underlined} indicate best and second best, respectively.
    }
    \begin{tabular}{ c | cccc | c }
        Model & $B \rightarrow G$ & $B \rightarrow M_o$ & $B \rightarrow M_r$ & $B \rightarrow V$ & Overall \\
        \hline
        SAN \shortcite{ref:ALLS20} &
            0.0602 ± 0.0414 &
            \textbf{0.0464 ± 0.0284} &
            0.0615 ± 0.0407 &
            0.0713 ± 0.0484 &
            0.0598 ± 0.0412	\\
        Ours - no SJS &
            \underline{0.0495 ± 0.0092} &
            0.0726 ± 0.0158 &
            0.0486 ± 0.0085 &
            0.0502 ± 0.0078 &
            0.0552 ± 0.0147 \\
        Ours - no BLC &
            0.0514 ± 0.0108 &
            0.0706 ± 0.0200 &
            \textbf{0.0308 ± 0.0107} &
            \textbf{0.0338 ± 0.0087} &
            \underline{0.0466 ± 0.0207} \\
        Ours &
            \textbf{0.0454 ± 0.0105} &
            \underline{0.0597 ± 0.0186} &
            \underline{0.0348 ± 0.0099} &
            \underline{0.0361 ± 0.0086} &
            \textbf{0.0440 ± 0.0160} \\
    \end{tabular}
    \label{tab:cross}
\end{table*}

    While approaches to motion retargeting such as skeleton-aware networks (SAN) \cite{ref:ALLS20} are specifically trained on a subset of the Mixamo dataset, ours requires a variation of skeleton topology in the training set and hence has been trained on a dataset that aggregates 4 different databases (see \cref{sec:Appendix:Data}) excluding Mixamo. To make the comparison more fair, we additionally fine-tuned our model on a few motion sequences from Mixamo (not used in the following evaluation) before evaluating our model on motion retargeting. The procedure (see \cref{sec:Appendix:FineTuning}) is roughly the same as the pre-training of our model, except that the amount of motion data is far smaller and its duration is much shorter.
    
    Finally, we observed a lower performance of our model on characters with body proportions very different from humans, and therefore detail our evaluation per character to emphasise the performance and limitations of our model depending on body proportions. The reason here is that we trained our model on motion data captured from human subjects, while in our evaluation two characters ($V$ and $M_o$) have body proportions that significantly differ from human ones.
    
	\paragraph{Self-Retargeting.}	
		\cref{tab:self} shows that our model performs well on all test characters for self-retargeting, even on non-human morphologies. Nevertheless, the retargeting error is significantly higher for Mousey compared to other characters. We notably outperform SAN \cite{ref:ALLS20}, with an average retargeting error normalised by character height at 0.009, which is equivalent to an error of 1.62~cm for a 1.8~m tall character, compared to 0.018 and 3.24~cm respectively for \cite{ref:ALLS20}.
	
	\paragraph{Intra-Structural Retargeting.}
		As shown in \cref{tab:intra}, the performance of our model on intra-structural retargeting is lower than SAN \cite{ref:ALLS20}. The reason is twofold: first, non-human morphologies are involved more often. In particular, the retargeting error when Mousey is either the source or target character is increased with respect to other characters, which points out that the ability of our model to generalise to morphologies very different from the ones in the training set is somehow limited. Second, for each skeleton topology, SAN \cite{ref:ALLS20} relies on dedicated encoder and decoder networks, which seems to help specialising on intra-structural retargeting. However, this might be at the expense of a latent space that could not be fully shared across skeleton topologies. The lower performance of our model on intra-structural retargeting suggests that some fine morphological features are less correctly transferred than with SAN, which we elaborate on and provide possible directions to improve on in \cref{sec:Conclusion}.
		
\begin{figure*}
	\centering
	\includegraphics[width=\textwidth]{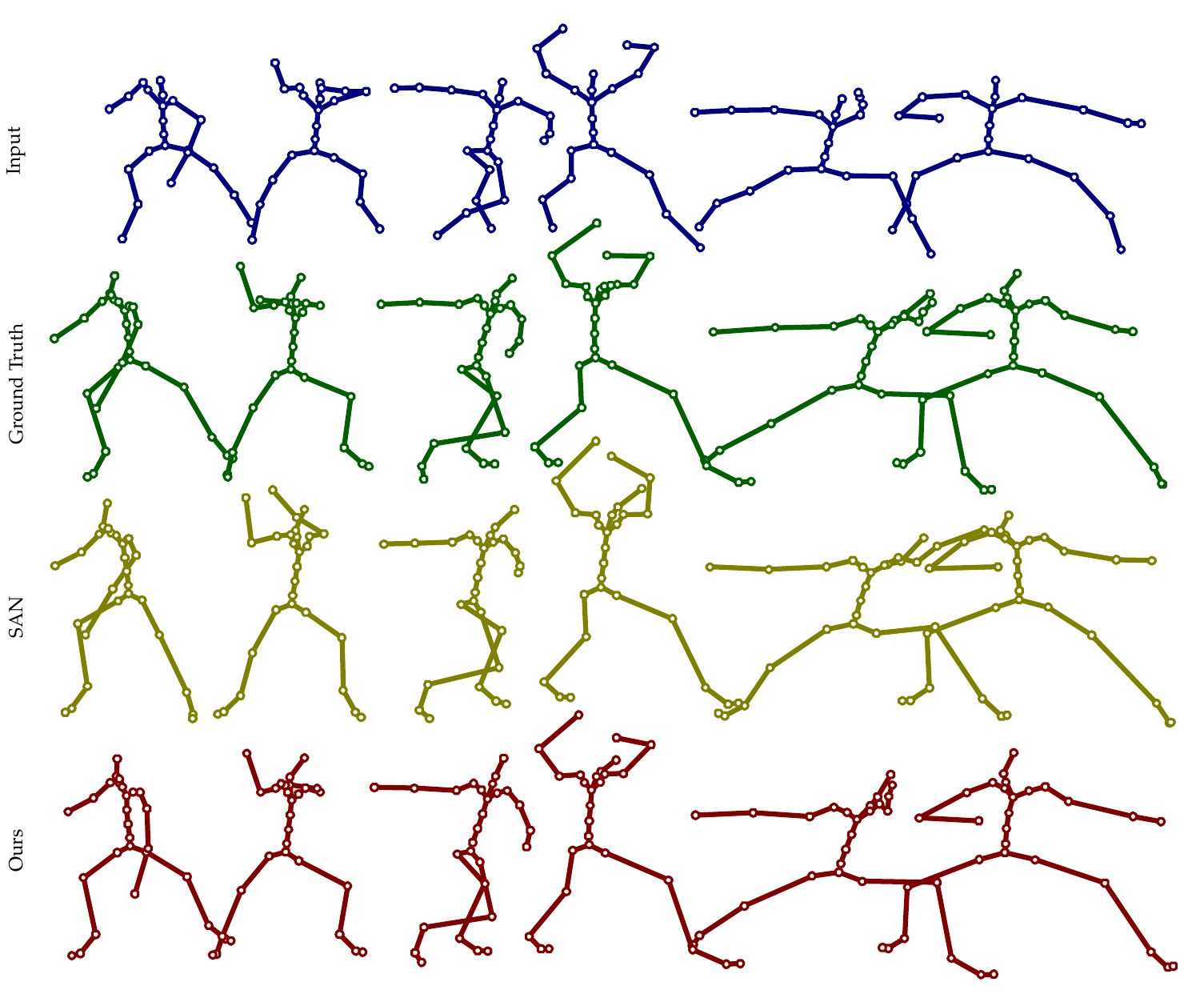}
    \caption{
		Visual result of cross-structural retargeting. In this example, a motion sequence (1st row) drawn from Mixamo, consisting in fighting moves, is retargeted from \emph{BigVegas} to \emph{Vampire} using either skeleton-aware networks \cite{ref:ALLS20} (SAN, 3rd row) or our model (Ours, 4th row). The corresponding sequence for character \emph{Vampire} (2nd row) is considered as ground truth when evaluating motion retargeting.
    }
    \label{fig:retargeting_m}
\end{figure*}

\begin{figure*}
	\centering
	\includegraphics[width=\textwidth]{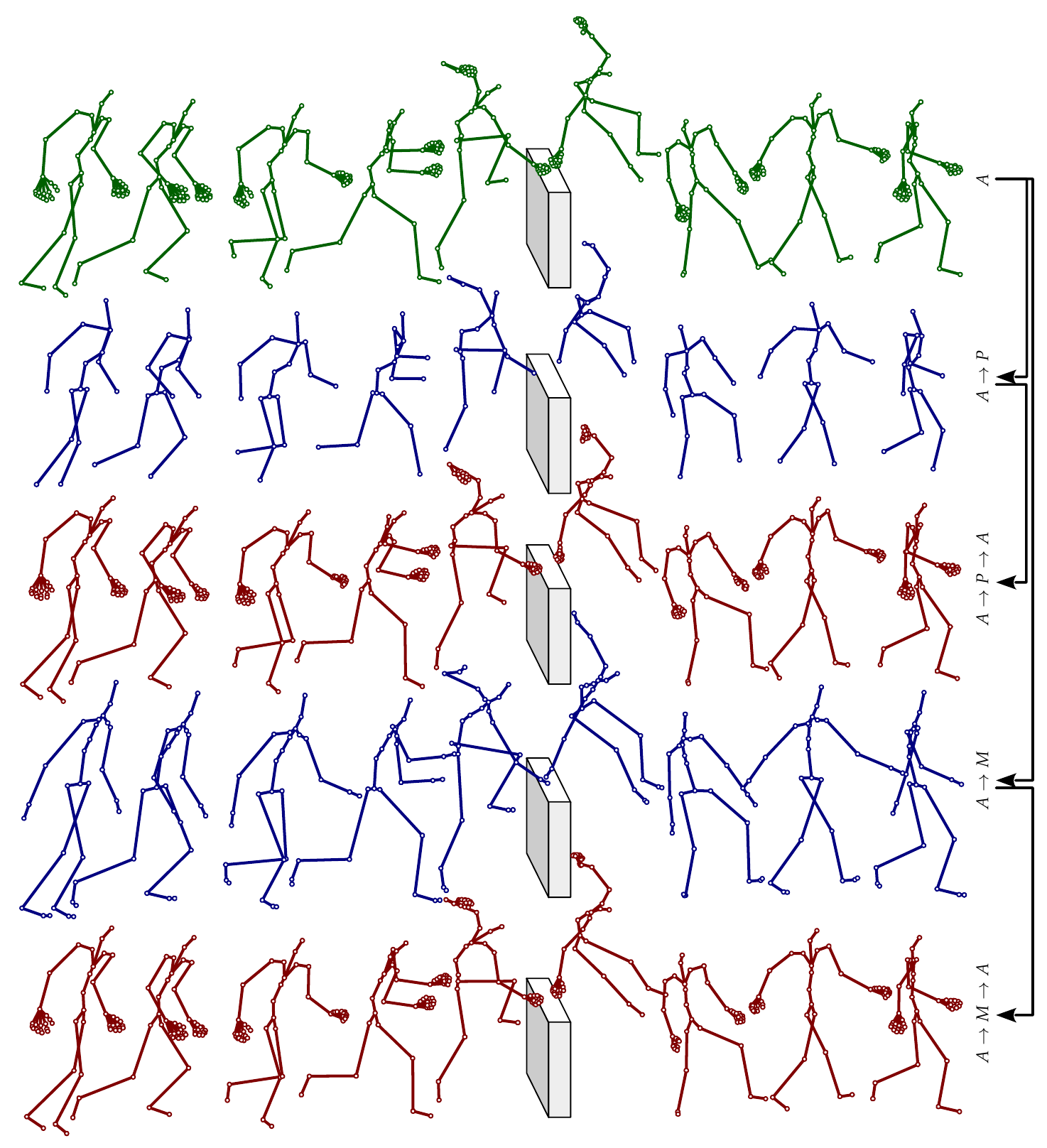}
    \caption{
		Visual examples of motion retargeting. A motion sequence (see 1st row, in green) with AMASS skeleton topology $A$ is retargeted to PSU-TMM100 and MPI-INF-3DHP topologies, noted $P$ and $M$, respectively (see 2nd and 4th rows, respectively, in blue). Then, both are retargeted back to AMASS topology (see 3rd and 5th rows, respectively, in red). Note that our model performs well on MPI-INF-3DHP topology, even though it has never seen it during training.
    }
    \label{fig:retargeting}
\end{figure*}

	\paragraph{Cross-Structural Retargeting.}
	    \cref{tab:cross} confirms our thoughts as our model outperforms SAN \cite{ref:ALLS20} on cross-structural retargeting, which is more difficult than intra-structural retargeting and for which it is critical that the latent space is well shared across topologies. Hence, the performance of our model suggests that it effectively captures abstract motion features, while the latent space of SAN \cite{ref:ALLS20} might not be as much shared among topologies as ours even though performing better on intra-structural retargeting. The only character for which SAN performs slightly better than our model is again for retargeting towards Mousey, i.e., to the morphology most different from the ones seen during training of our model. Finally, \cref{fig:retargeting_m,fig:retargeting} show visual results of cross-structural motion retargeting on sample motion sequences from Mixamo and our validation set (see \cref{sec:Appendix:Data}), respectively.
    
	\paragraph{Ablative Study.}
	    \cref{tab:reconstruction,tab:self,tab:intra,tab:cross} also confirm that the stochastic joint subsampling mechanism and bone length consistency loss term are effective, as these variants \textit{Ours~-~no~SJS} and \textit{Ours~-~no~BLC} are outperformed by our reference model. However, both variants give competitive results and are even slightly better in specific cases. Our reference model still provides the best performance on average.
	    
\subsection{Motion Denoising}\label{sec:EvalMotionDenoising}
	In this section, we demonstrate that our model also has the ability to effectively clean noisy motion sequences by projecting them into the learnt latent space of human motion sequences. This suggests that the underlying deep motion representation is well structured, a consideration of particular relevance for the retargeting task. To this end, we pick clean motion sequences from our validation set and purposely add Gaussian noise on joint positions to obtain corresponding noisy motion sequences. Then we simply encode and decode these noisy motion sequences and compare them to the corresponding original clean motion sequences. We follow this procedure for an entire range of noise levels, controlled by the standard deviation of offset coordinates (sampled from a centred Gaussian) that are added to coordinates of joint positions.
	
\begin{figure}
    \small
    \centering
	\includegraphics[width=\linewidth]{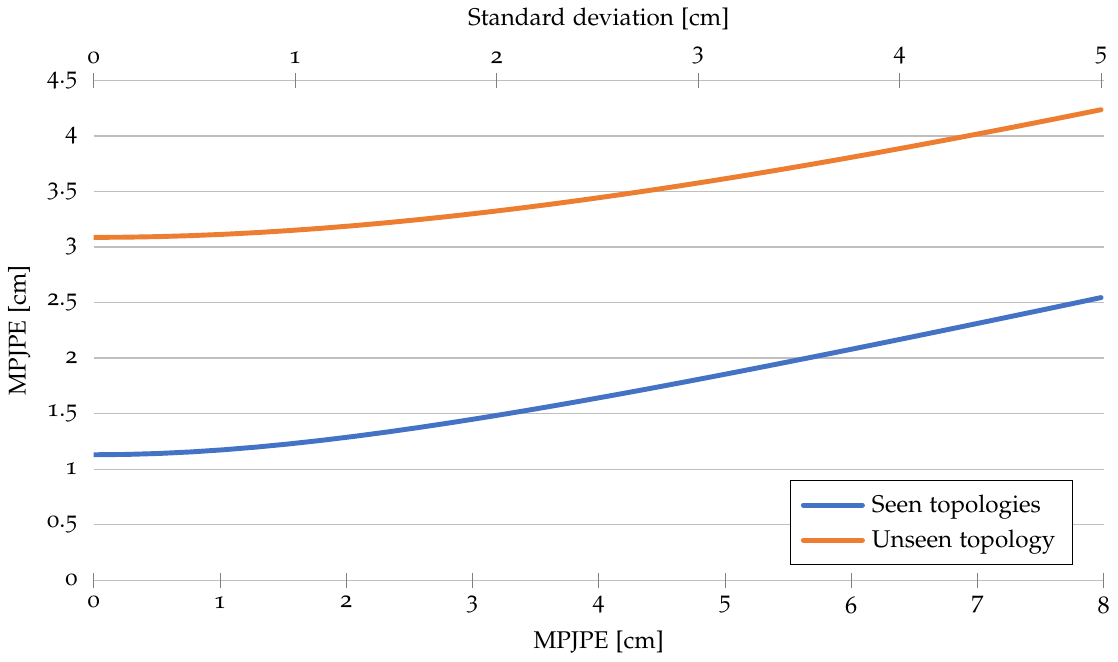}
    \caption{
		Quantitative evaluation of our model on motion denoising. Gaussian noise is purposely added to clean motion sequences from our validation set. The resulting noised motion sequences are then encoded and decoded using our model to perform denoising. For an entire range of levels of noise, positional error on denoised sequences is plotted vertically against the positional error observed in noised sequences, horizontally. The top axis indicates the standard deviation of the different levels of noise, while the bottom axis gives the MPJPE observed in corresponding noised sequences, which is proportional to the standard deviation. Results are averaged over seen and unseen skeleton topologies. Flatter curves correspond to higher denoising ability.
	}
    \label{fig:denoising}
\end{figure}

	In \cref{fig:denoising}, we vertically plot the MPJPE of denoised motion sequences with respect to clean motion sequences, against the levels of added noise horizontally. The MPJPE is computed for all motion sequences from our validation set and averaged over seen and unseen skeleton topologies. Without noise (at $x=0$ in \cref{fig:denoising}), we obtain the representation errors found when evaluating the representation accuracy of our model (see \cref{tab:reconstruction}). Then, positional errors observed in denoised motion sequences increase with noise but only slowly, indicating that our model effectively removes noise. The right-hand side of \cref{fig:denoising} demonstrates the ability of our model to reduce the noise level as the output positional error is lower than the input positional error observed in noisy motion sequences. Animated visual examples can be found in the supplementary video; they provide more meaningful illustrations of the ability of our model to perform motion denoising.

\begin{figure}
    \small
    \centering
	\includegraphics[width=\linewidth]{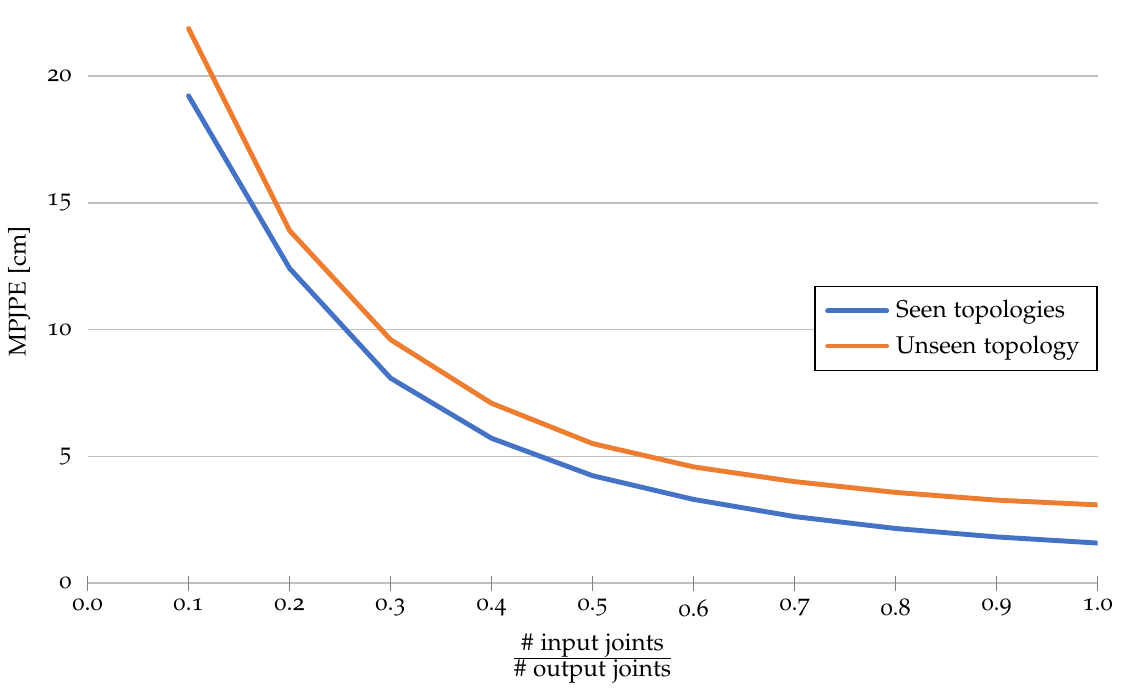}
    \caption{
		Quantitative evaluation of our model on joint upsampling. Here, we purposely discard random joints from motion sequences drawn from our validation set and then reconstruct missing joints using our model. To do so, encoding is conditioned on the subsampled version of skeleton templates while full skeleton templates condition the decoding. We then plot the reconstruction error measured by the MPJPE vertically against the proportion of input joints plotted horizontally.
	}
    \label{fig:upsampling}
\end{figure}

\subsection{Joint Upsampling}\label{sec:EvalJointUpsampling}
    Finally, we demonstrate in this section that our model can also be applied to the (spatial) upsampling of joints in motion sequences. We perform joint upsampling by encoding a low-resolution motion sequence into our topology-agnostic motion representation and then running the decoder conditioned on an upsampled version of the encoding skeleton template. In other words, this is a particular case of retargeting where the source skeleton consists of a subset of the joints of the target skeleton.
    
\begin{figure*}
	\centering
	\includegraphics[width=\textwidth]{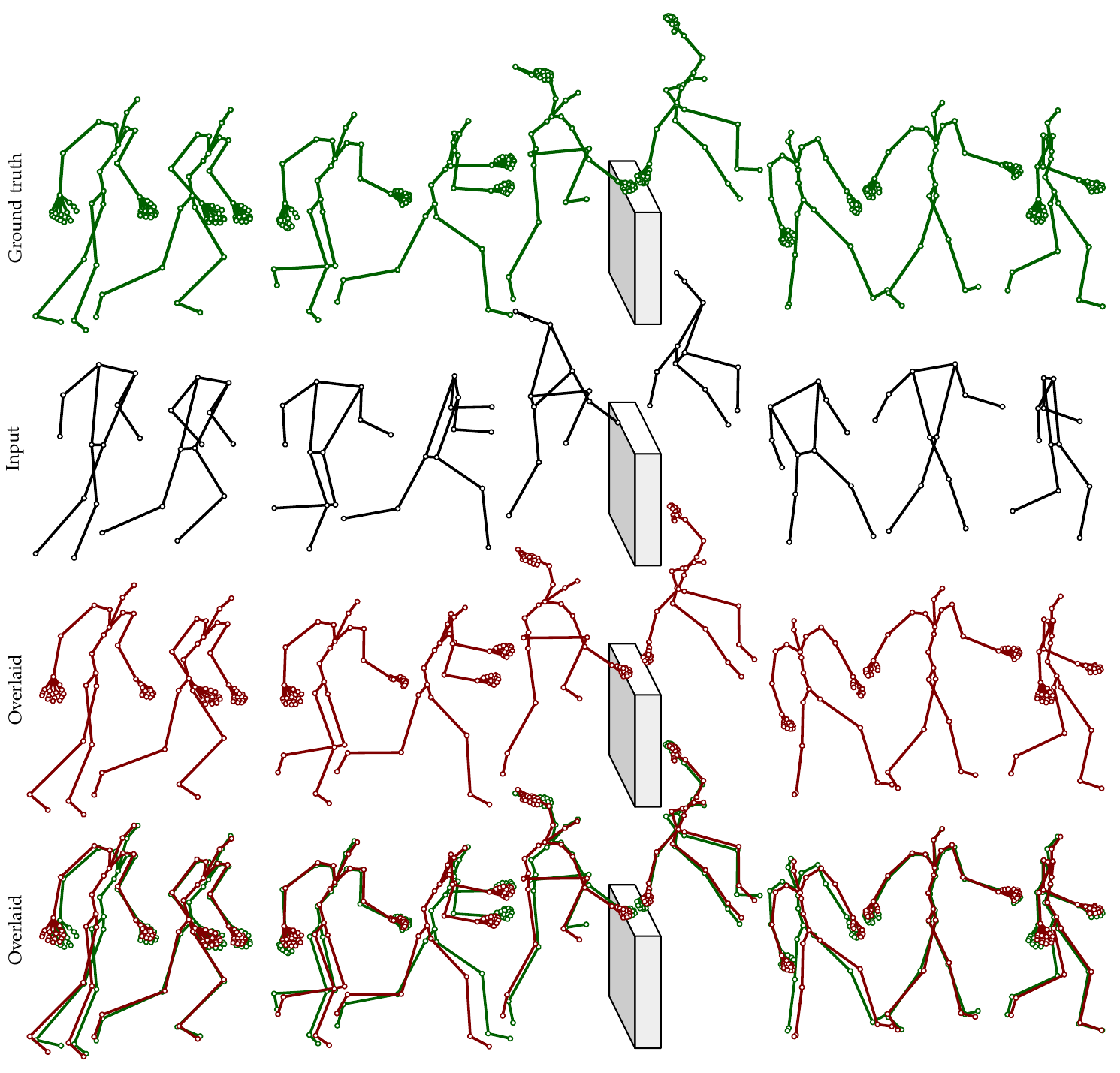}
    \caption{
		Illustration of joint upsampling performed using our model. In this example, a ground truth test motion sequence (top row, green) is purposely downsampled by keeping only ankles, knees, hips, shoulders, elbows and wrists (second row, black). This subsampled motion sequence is then upsampled back to the original joint resolution using our model (\ie retargeting from the low-resolution skeleton to the high-resolution skeleton), as depicted in third row, in red. The bottom row shows both ground truth and upsampled motion sequences overlaid to emphasise errors. In this particular example, 23.1\% (54.6\% if not considering fingers) of the joints are given in input to our model, which perform upsampling with a resulting MPJPE of 4.17 cm (2.85 cm if not considering finger joints).
    }
    \label{fig:upsvis}
\end{figure*}

    We quantify the ability of our model to perform joint upsampling on our test set (see \cref{sec:Appendix:Data}) by randomly selecting joints in test motion sequences (others are discarded), and then by upsampling them back to the original joint resolution using our model as described above. Finger joints are not considered (neither in inputs nor in outputs) to homogenise joint proportions across the different skeleton topologies and hence minimise biases. \cref{fig:upsampling} shows the resulting joint upsampling error, measured by the MPJPE, for different proportions of joints present in the subsampled inputs. Our model performs quite well up to a joint subsampling factor of $0.5$, with an error of about 5 cm (slightly below on seen skeleton topologies and slightly above on unseen topology). Animated visual results provided in the supplementary video show that these errors are well distributed and that resulting upsampled motion sequences are perceptually quite plausible. Moreover, \cref{fig:upsvis} shows a visual example of joint upsampling using our model.

    \section{Conclusion} \label{sec:Conclusion}
        In this article, we have presented a novel approach for motion retargeting that leverages a transformer-based autoencoder to learn an abstract representation of human motion that is agnostic to skeleton topology and morphology. Our model has demonstrated the ability to encode and decode motion sequences with variable skeleton topologies. It achieves state-of-the-art performance on motion retargeting and extends the scope of retargeting beyond homeomorphic skeleton topologies seen during training. Our approach addresses the limitations of existing methods for motion retargeting, which often focus solely on morphological variations and are limited to known skeleton topologies. By training on a diverse dataset gathered from multiple existing databases, our model successfully abstracts motion features from morphological and topological features, enabling the retargeting of motion across different characters. Beyond motion retargeting, the abstract representation of motion provides a convenient space to combine motion data from different sources, enabling the integration of small specialised motion datasets with larger general motion datasets for improved performance and generalisation ability. As shown in our evaluation as well as in the animated visual results provided in the supplementary video, the proposed approach also performs well on other tasks such as motion denoising and joint upsampling. This demonstrates that our latent space motion representation effectively captures the human motion information in skeleton sequences.

It is tempting to think that our model can generalise to any morphology. However, our evaluation shows that our model learnt on human characters is more effective on human-like morphologies than on caricatural morphologies. We also show that our model is less accurate on intra-structural retargeting, \ie with source and target characters having the same topology. The reason might be that the skeleton templates lack the complexity needed to encode fine morphological variations.

Skeleton templates that we represented with joint positions have no mechanism to specify correspondences. Thus, a possible direction of future work is to refine skeleton templates and the conditioning mechanism. Moreover, the success of the proposed transformer-based framework for motion retargeting is promising in many other fields of character animation. A similar network could be trained to predict a few frames in the near future from a motion sequence in the same way models in natural language processing predict upcoming words. Such a model could then be leveraged in various applications, such as in motion synthesis or in online systems to reduce latency.

In conclusion, the proposed abstract representation and transformer-based autoencoder provide a promising framework for motion retargeting and related tasks. The ability to transfer motion dynamics across characters with varying morphologies and topologies opens up new possibilities for character animation and motion editing, enhancing interoperability and flexibility in animation pipelines.

    
    % bibliography
    \bibliographystyle{ACM-Reference-Format}
    \bibliography{bibliography.bib}

    \appendix
    \section{Implementation details.} \label{sec:Appendix}
    In this section we give implementation details necessary for reproducibility. Please also refer to our code and pre-trained model, which we make available at \ImplementationLink.
    
	\subsection{Data} \label{sec:Appendix:Data}
		To train and evaluate our model, we split our dataset into a training set and a validation set. The latter is constituted of all chunks from MPI-INF-3DHP~\cite{ref:MRCF17}, for validation on an unseen skeleton topology, as well as about 10\% of chunks from other sub-datasets for validation on known skeleton topologies. This results in about 114,275 chunks with topologies seen during training, and 26,249 chunks with MPI-INF-3DHP topology unseen during training.
 
    \subsection{Skeleton Templates} \label{sec:Appendix:Templates}
	    As explained in \cref{sec:TemplateEncoding}, we manually predefined a generic neutral pose for each data source as illustrated in \cref{fig:topologies}. Then, we obtain the skeleton template for a given motion sequence by scaling the bone lengths of the corresponding generic neutral pose to match the bone lengths observed in the given motion sequence. Since some of our sources of motion data use a positional pose representation, the bone lengths sometimes vary a little over time. Consequently, we use the temporal median of bone lengths of a given motion sequence as observed bone lengths.

    \subsection{Architecture} \label{sec:Appendix:Architecture}
        As illustrated in \cref{fig:encoder}, our encoder begins with two joint-wise embedding modules: a 3-layer MLP is used to embed skeleton templates (16, 32 and 64 output features) while motion sequences are embedded by a CNN made of 4 \D convolutional layers (3, 16, 32 and 64 output features) followed by 3 linear layers (64 output features each). Each layer of both modules is followed by an exponential linear unit (ELU). Then, the core of our encoder is made of 4 transformer blocks with 8 heads and 128 features each. Finally, intermediate and final linear layers (see \cref{fig:encoder}) have both 128 output features. Similarly to our encoder, templates and latent codes are also embedded at the beginning of our decoder (see \cref{fig:decoder}). Skeleton templates are embedded by a 4-layer MLP (16, 32, 64 and 128 output features) with rectified linear units (ReLUs) after each layer except the last. Latent codes are embedded by a CNN made of 3 \D convolutional layers followed by a linear layer (128 output features and ELU activation each). Then, the core of our decoder is also made of 4 transformer blocks with 8 heads and 128 features each. Moreover, a 2-layer MLP with 128 output features and ReLU activation is associated to each block to further independently process latent features before modulating input tokens (see \cref{sec:Decoder}). After transformer blocks, a final joint-wise CNN is used to reconstruct motion sequences. It is made of 3 \D convolutional layers (128 output features) followed by 3 linear layers (128, 128 and 3 output features), with an ELU activation after each layer except the last. As explained in \cref{sec:XCA}, transformer blocks in both our encoder and decoder leverage XCA. In total, our model has 2,475,511 learnable parameters.

	\subsection{Stochastic Joint Subsampling}  \label{sec:Appendix:SJS}
	    As seen in \cref{sec:Losses}, we apply a stochastic joint subsampling mechanism during training, which discards random sets of joints. In practice, we discard each joint independently with probability $p=0.1$ for major joints (ankles, hips, knees, shoulders, elbows and wrists) that are common to all skeleton topologies of our dataset, and with probability $p=0.5$ for other joints. Accordingly, our loss function is computed over the joints of the decoder skeleton template (reconstruction loss) and the corresponding bones (bone lengths consistency loss).

	\subsection{Training} \label{sec:Appendix:Training}
		Our implementation is written in Python and relies on PyTorch. Training was performed on an NVidia Ampere A100 GPU while other results were obtained either on an NVidia GeForce RTX 2060 GPU or on CPU. We trained our model for $10^{7}$~iterations (about 385~epochs and 26.5~days) using the Adam optimisation algorithm~\cite{ref:KiBa14} with a mini-batch size of 32 and hyperparameters $\beta_{1} = 0.9$, $\beta_{2} = 0.999$. Moreover, we used a triangular cyclic learning rate~\cite{ref:Smit17} with a period of 10~epochs and lower and upper bounds set at $10^{-4}$ and $10^{-5}$, respectively. Finally we set the weight of our temporal bone lengths consistency loss to ${\lambda_{blc} = 0.5}$ (see \cref{sec:Losses}).
	
	\subsection{Fine-Tuning on Mixamo} \label{sec:Appendix:FineTuning}
    	To have a fair evaluation of our model (which has not been trained on Mixamo) against skeleton-aware networks \cite{ref:ALLS20}, we additionally fine-tuned our model on a few motion sequences from Mixamo. The procedure is the same as the training of the model, except that the amount of motion data is far smaller (about 1 hour vs 65 hours for training) and its duration is much shorter (less than 2 hours vs 26.5~days). We also used a constant learning rate set at $5\cdot10^{-5}$ instead of the cyclic learning rate. Finally, since the pelvises of skeleton templates are aligned (see \cref{sec:TemplateEncoding}), our model aligns the global trajectory of output motion sequences with respect to input motion sequences around the same region of the body. However, when performing motion retargeting on source and target characters of different scales, the resulting global trajectory might not be consistent with the target character scale. To fix this issue, we further scale the global trajectory of output motion sequences when performing motion retargeting based on the leg length ratio of source and target characters, as typically performed in the literature \cite{ref:KuMA05, ref:HRED08}.

\end{document}